\documentclass[letterpaper]{article} 
\usepackage[preprint]{aaai2027}  
\usepackage[hyphens]{url}  
\usepackage{graphicx} 
\usepackage{natbib}  
\usepackage{caption} 
\usepackage{booktabs}
\usepackage{amsmath}
\usepackage{amssymb}


\newcommand{\repositoryurl}[1]{%
\leavevmode\pdfstartlink attr{/Border [0 0 0]} user{/Subtype /Link /A << /S /URI /URI (#1) >>}%
\url{#1}\pdfendlink}
\newcommand{\pp}{\,pp}

\title{Same Contacts, Different Rooms:\\Meeting Format Shapes LLM Group Influence Beyond Pairwise Contact Counts}
\author{Zhexi Gu\textsuperscript{1}, Shusheng Cao\textsuperscript{2}, Runping Zhu\textsuperscript{3}}
\affiliations{
\textsuperscript{1}School of Information and Library Science, University of North Carolina at Chapel Hill, USA\\
\textsuperscript{2}School of Geographic and Biologic Information,\\
Nanjing University of Posts and Telecommunications, China\\
\textsuperscript{3}Public Relations and Advertising, Department of Communication,\\
Faculty of Humanities and Social Sciences, Beijing Normal-Hong Kong Baptist University, China\\
zhexigu@unc.edu, caoss@njupt.edu.cn, serenerpzhu@bnbu.edu.cn
}

\begin{document}

\maketitle

\begin{abstract}
Simulations of LLM agent collectives usually summarise interaction as a pairwise contact network. We test that summary with a contact-matched intervention: the members, their initial stances and every pairwise contact count stay fixed while the meetings are reorganised, and each member gives a one-step support readout after reading the arguments of everyone present. Across six open-weight LLMs and 14 debate topics, merging triads into five-person meetings (\emph{resize}) moves group support more than reshuffling who meets in triads (\emph{regroup}), although both move the same 60 dyadic exposures. The root-mean-square change is 4.36 versus 2.59 percentage points (pp), larger for resize in all six models and in 81 of 84 model--topic cells. Resizing also has a consistent direction: it amplifies the initial majority of the merged members by 1.72\pp, in every model, whereas regrouping has no direction ($\le 0.03$\pp). A contact network cannot represent regrouping at all, yet pairwise rules applied meeting by meeting explain up to 76\% of its effect, and with identical weights they raise resize $R^2$ by 0.51--0.58 for two models. We show that the gap between the two placements is an exact Jensen gap of the sigmoid readout, bounded by how much a member's predicted logits vary across meetings; this dispersion tracks the gap with log--log $r=0.97$ and tells a modeller when contact counts are a safe summary. Meeting format is a first-class variable for LLM social simulation.
\end{abstract}

\section{Introduction}

Simulations of LLM agent collectives need a model of who influences whom. The usual answer is a network: agents are nodes, and an edge records how often two agents interacted \citep{castellano2009statistical,chuang2024simulating}. That representation discards the \emph{format} of each interaction: whether two agents met in a small group or a large one, and who else was in the room. Network science shows that group interactions can generate dynamics no pairwise network reproduces \citep{battiston2020networks,iacopini2019simplicial,neuhauser2020multibody}, and also that some group models reduce exactly to weighted pairwise networks \citep{llabres2026reducibility}. Which case holds for LLM agents is an empirical question, and an attribution question: when a collective outcome changes, is the cause the pattern of pairwise contacts or the way the meetings were organised?

\begin{figure*}[t]
\centering
\includegraphics[width=\textwidth]{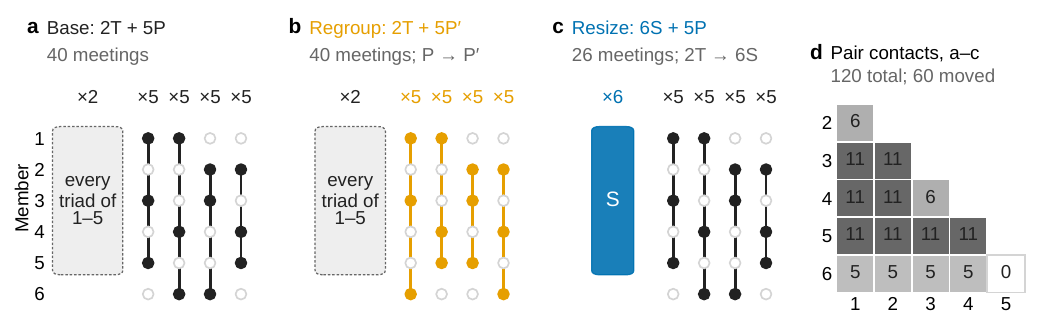}
\caption{Contact-matched schedules for six members (rows). Filled dots mark the members of one triad; ``$\times k$'' is its number of copies. (a)~Base: two copies of every triad among members 1--5 (2T) plus five copies of each of four further triads P. (b)~Regroup replaces P with four other triads P$'$ (orange). (c)~Resize replaces 2T with six five-person meetings S of members 1--5. (d)~All three schedules share this pair-contact matrix (120 dyadic contacts); each intervention removes and adds 60. Meeting counts are 40, 40 and 26; members 1--5 attend 22 meetings in base and regroup and 16 in resize, and member 6 attends 10 in all three.}
\label{fig:design}
\end{figure*}

We answer it with an intervention that holds pairwise contacts fixed. Six LLM members with fixed initial stances meet under one of three schedules (Figure~\ref{fig:design}) that produce the same pair-contact matrix. \emph{Regroup} changes which members meet in triads; \emph{resize} merges the triads among five members into five-person meetings. Each moves the same 60 dyadic exposures, so any difference in group support between schedules is caused by meeting format rather than by who met whom or how often. In each meeting, members read the arguments of everyone present and report their support. The outcome is the mean one-step response a schedule elicits, which is the quantity a sequential simulation would carry into its next state.

The intervention reveals a format effect in every model we test, and a specific one: larger meetings amplify the initial majority of their members. We then ask which pairwise models predict it. A static contact network cannot represent regrouping, because base and regroup share the same network. The same pairwise rules applied meeting by meeting can, and they also predict resizing much better for two of six models. We show that the difference between the two placements of one rule is an exact Jensen gap of the sigmoid readout, controlled by the dispersion of each member's predicted logits across meetings. This turns the choice between a contact network and a meeting-level model into a quantity that can be computed in advance.

Our contributions are:
\begin{itemize}
\item \textbf{A contact-matched format intervention.} A schedule design that fixes members, initial stances and every pairwise contact count while changing meeting format, run on six open-weight LLMs and 14 topics (161,280 matched configurations). Resizing moves one-step group support more than regrouping in every model and systematically amplifies the majority of the merged members.
\item \textbf{Weight form separated from aggregation placement.} Five low-parameter pairwise predictors, transferred from an independent micro-experiment without fitting on group outcomes, with aggregation placement isolated by applying identical weights meeting by meeting and on the contact network. Placement alone adds 0.51--0.58 $R^2$ for two models and recovers up to 76\% of the regroup effect that the network misses.
\item \textbf{An exact account of aggregation error.} The gap between meeting-wise and network predictions of the same rule is a Jensen gap bounded by the variance of meeting-conditional logits. It predicts where the two placements disagree and shows that identical effective edge weights do not imply identical predictions.
\end{itemize}

\section{Related Work}

\paragraph{Group size and majorities in LLM collectives.}
Multi-agent debate improves reasoning \citep{du2024improving}, and accuracy scales with the number of sampled agents \citep{li2024more}. Treated as influence processes, LLM groups show conformity to majorities whose strength depends on majority size \citep{zhang2024exploring,weng2025conformity}, can converge on a shared majority view even when it is wrong \citep{estornell2024multi}, form conventions that committed minorities can overturn \citep{ashery2025emergent}, and follow majorities with a force that weakens with group size \citep{demarzo2026majority}. LLM conformity has been fitted with social impact theory \citep{bellina2026prompt,latane1981psychology}, and the share of adversaries, rather than their number, predicts defection in deliberation \citep{wu2026adversarial}. Human work has the same themes \citep{asch1956studies,bond2005group,centola2018experimental}. These studies vary group size or composition, which also changes how often each pair of agents interacts. We hold every pair-contact count fixed and change only how contacts are packaged into meetings.

\paragraph{Higher-order interactions and pairwise reduction.}
Hypergraph and simplicial models capture interactions among more than two units \citep{battiston2020networks,battiston2021physics}. Group contagion produces discontinuous transitions that pairwise networks miss \citep{iacopini2019simplicial}. Genuine multibody effects require a nonlinear interaction function \citep{neuhauser2020multibody,sahasrabuddhe2021modelling}, and consensus on temporal hypergraphs can differ sharply from consensus on their static aggregate \citep{neuhauser2021temporal,holme2012temporal}. Conversely, social impact models on hypergraphs map exactly onto weighted pairwise networks, with state-dependent weights when impact is nonlinear \citep{llabres2026reducibility}. Classical opinion dynamics work with pairwise rules \citep{castellano2009statistical,friedkin1990social}. LLM agents come with no known update rule, so we test reducibility empirically, and we separate two choices that a reduction bundles together: the form of the weights and the level at which they are aggregated.

\paragraph{LLM social simulation and measurement.}
Generative agents \citep{park2023generative}, silicon samples \citep{argyle2023out} and simulated subject studies \citep{aher2023using} established LLMs as simulation tools, and validation is now the central concern \citep{anthis2025llm,larooij2025validation}. We contribute a controlled intervention on the behaviour of LLM collectives themselves. Our readout follows option log-probability methods \citep{santurkar2023whose}, which can carry label biases and diverge from generated text \citep{dominguezolmedo2024questioning,wang2024answer}; every contrast we report is therefore a within-model difference between schedules with identical prompts and options, and the main contrast is reproduced with a full-answer likelihood readout.

\section{Design and Measurement}
\label{sec:design}

\paragraph{Members and topics.}
Topics are debate motions from ArgKP-2021 \citep{barhaim2020arguments,friedman2021overview}. For each topic we form a pool of 15 members. Each member holds one human-written argument on each side, and its initial stance determines the side it argues. All members in a run are instances of one base model and differ only in stance and argument. The main study uses six small open-weight models (Appendix Table~\ref{tab:models}) and 14 topics (Appendix~\ref{app:design}). A configuration fixes a five-member core (members 1--5; three disjoint cores per topic), a sixth member drawn from the ten pool members outside the core, and one of the $2^6=64$ initial stance patterns, giving 1,920 configurations per model and topic.

\paragraph{Schedules.}
Base is $2\mathrm{T}+5\mathrm{P}$: two copies of each of the ten triads among members 1--5, plus five copies of each of four further triads P, two of which include member 6 (Figure~\ref{fig:design}). Regroup, $2\mathrm{T}+5\mathrm{P}'$, replaces P with four triads P$'$ that cover the same pairs. Resize, $6\mathrm{S}+5\mathrm{P}$, replaces 2T with six five-person meetings of members 1--5. The three schedules have identical pair-contact matrices (Figure~\ref{fig:design}d), and each intervention removes and adds 60 dyadic exposures. Resize also has fewer meetings (26 instead of 40), so the contrast is a format intervention at a fixed pair-contact budget rather than at fixed meeting count or token cost.

\paragraph{Readout and group support.}
In each meeting, every attending member sees the arguments of the other attendees and answers a binary support/oppose question. We read the logit difference $\ell_{ie}$ between the two option tokens and set $p_{ie}=\sigma(\ell_{ie})$. Every meeting starts from the members' initial stances, so the readout is a one-step response to the people present. Group support under a schedule $\mathcal{M}$ is
\begin{equation}
Q(\mathcal{M})=\frac{1}{6}\sum_{i=1}^{6}\frac{1}{m_i}\sum_{e\in\mathcal{M}:\,i\in e} p_{ie},
\label{eq:Q}
\end{equation}
where $m_i$ is the number of meetings member $i$ attends (\emph{person} weighting). \emph{Incidence} weighting instead averages over all member--meeting events.

\paragraph{Observables.}
For each configuration $c$, the resize effect is $\Delta^{\mathrm{res}}_c=Q(\mathrm{resize})-Q(\mathrm{base})$ and the regroup effect is $\Delta^{\mathrm{reg}}_c=Q(\mathrm{regroup})-Q(\mathrm{base})$. Magnitude is the model-level root mean square, $\mathrm{RMS}=100\,(\mathrm{mean}_t\,\mathrm{mean}_c\,\Delta_c^2)^{1/2}$ in pp, where $t$ indexes topics. The pooled difference averages the six model-level RMS differences with equal weights. Direction is the majority-directed shift $D=100\,\mathrm{mean}_c\,s_c\Delta_c$, with $s_c=+1$ if at least three of members 1--5 start in support and $s_c=-1$ otherwise.

\paragraph{Inference.}
Intervals are percentile 95\% intervals from a topic-cluster bootstrap (14 topics, 20,000 draws). The six models are treated as fixed, and pooled estimates resample topics jointly across models.

\section{Pairwise Predictors}
\label{sec:predictors}

\paragraph{Weight form.}
Let $a_i$ be member $i$'s logit with no peers present and $v_{ij}$ the change in that logit when only $j$ is present. Both are measured for each configuration. For member $i$ in meeting $e$ with peers $N=e\setminus\{i\}$ and $n=|N|$,
\begin{equation}
z_{ie}=a_i+\sum_{j\in N}\omega\, v_{ij},
\label{eq:z}
\end{equation}
with $\omega=1$ (\textbf{Sum}), $\omega=1/n$ (\textbf{Mean}), $\omega=w_n$ (\textbf{Size} weights), or separate weights $w^{\mathrm{opp}}_{n,o}$ and $w^{\mathrm{ally}}_{n,o}$ for opponents and allies of $i$ when $o$ of the $n$ peers are opponents (\textbf{Stance} weights). Size and stance weights come from a separate micro-experiment in which a focal member faces 0--6 opponents and 0--3 allies (64 items, 19 topics; Appendix~\ref{app:micro}), fitted with the target topic excluded (58--64 training items). Size weights are fitted by least squares through the origin; stance weights are ridge-shrunk toward the size weights with penalty $0.1\,\mathrm{tr}(X^\top X)/k$ for $k$ columns. No parameter is fitted on group outcomes, so every group-level prediction below is out of sample.

\paragraph{Aggregation placement.}
The \emph{meeting-wise} prediction applies Eq.~\ref{eq:Q} to $\hat p_{ie}=\sigma(z_{ie})$. The \emph{static network} prediction averages logits first and applies the sigmoid once:
\begin{equation}
\hat Q^{\mathrm{net}}=\frac{1}{6}\sum_i \sigma(\bar z_i),\qquad \bar z_i=\frac{1}{m_i}\sum_{e\ni i} z_{ie}.
\label{eq:net}
\end{equation}
With Size weights, $\bar z_i=a_i+\sum_j W_{ij}v_{ij}$ with $W_{ij}=m_i^{-1}\sum_{e\ni i,j}w_{|e|-1}$, a weighted contact network. When all meetings are triads, $W_{ij}=w_2c_{ij}/m_i$ depends on the schedule only through the contact counts $c_{ij}$. Base and regroup share every $c_{ij}$ and $m_i$, so the network predicts a regroup effect of exactly zero. Comparing Size meeting-wise with Size on the network isolates placement; comparing Size with Mean, and Stance with Size, isolates weight form.

\paragraph{Scoring.}
We use the contrast $R^2=1-\sum_t\mathrm{MSE}_t/\sum_t\mathrm{MS}_t$, where $\mathrm{MS}_t$ is the mean squared observed effect in topic $t$ and $\mathrm{MSE}_t$ the mean squared prediction error. $R^2=0$ is the predictor of no change, and $R^2$ is the share of the effect's second moment that a rule explains. Paired gains are $R^2$ differences against the same denominator.

\section{Results}

\begin{table*}[t]
\centering
\small
\setlength{\tabcolsep}{4.2pt}
\begin{tabular}{lcccccrccc}
\toprule
 & \multicolumn{5}{c}{Magnitude: RMS change (pp)} & & \multicolumn{3}{c}{Majority-directed shift $D$ (pp)} \\
\cmidrule(lr){2-6}\cmidrule(lr){7-10}
Model & Resize & Regroup & Resize $-$ regroup [95\% CI] & Ratio & Cells & Resize [95\% CI] & Regroup & Size w. & Network \\
\midrule
Granite4.2-3B & 5.61 & 4.38 & 1.23 [0.67, 1.83] & 1.28 & 13/14 & 2.85 [2.25, 3.48] & 0.00 & 2.26 & 0.62 \\
LFM2.5-1.2B & 3.09 & 2.30 & 0.79 [0.55, 1.02] & 1.34 & 13/14 & 0.93 [0.58, 1.24] & 0.00 & 1.11 & $-$0.16 \\
Gemma4-E2B & 6.70 & 3.87 & 2.83 [1.94, 3.65] & 1.73 & 13/14 & 2.27 [1.68, 2.76] & 0.00 & 0.68 & 0.25 \\
MiniCPM5-2B & 5.33 & 2.39 & 2.94 [2.15, 3.61] & 2.23 & 14/14 & 2.74 [2.10, 3.36] & 0.03 & $-$0.09 & $-$0.11 \\
Phi4-mini & 3.22 & 1.39 & 1.82 [1.28, 2.38] & 2.31 & 14/14 & 0.90 [0.44, 1.40] & 0.00 & $-$0.01 & $-$0.04 \\
Qwen3.5-4B & 2.23 & 1.20 & 1.04 [0.70, 1.35] & 1.87 & 14/14 & 0.61 [0.43, 0.81] & $-$0.01 & $-$0.04 & $-$0.09 \\
\midrule
All six & 4.36 & 2.59 & 1.77 [1.50, 2.03] & 1.28--2.31 & 81/84 & 1.72 [1.41, 2.02] & 0.00 & 0.65 & 0.08 \\
 
\bottomrule
\end{tabular}
\caption{Meeting-format effects on one-step support with pair contacts fixed (14 topics, 1,920 configurations each, person weighting). Cells: model--topic cells in which resize has the larger RMS. The last two columns give the majority-directed shift that the meeting-wise Size-weight rule and the static network rule predict for resize. All-six row: equal-weight model average with jointly resampled topics; Ratio there gives the range across models.}
\label{tab:effects}
\end{table*}

\begin{figure*}[t]
\centering
\includegraphics[width=\textwidth]{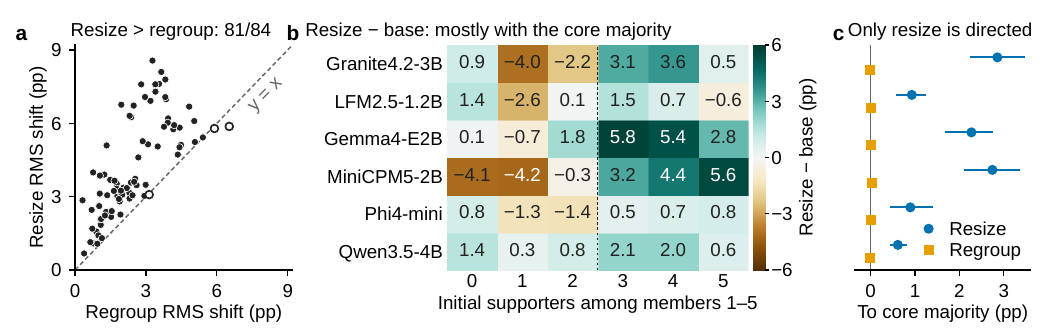}
\caption{Meeting format changes one-step group support at fixed pair contacts. (a)~RMS change per model--topic cell ($n=84$; 1,920 configurations per cell); filled points: resize larger, open points: regroup larger. (b)~Signed resize effect (pp), averaged over topics, by the number of initial supporters among members 1--5; the dashed line marks where the core majority switches from oppose to support. (c)~Majority-directed shift $D$ with 95\% topic-cluster intervals, rows as in (b). Regroup estimates are at most 0.03\pp\ in every model (its signed effects by core composition all lie within $\pm0.11$\pp), so only resize has a systematic direction.}
\label{fig:effect}
\end{figure*}

\subsection{Resizing moves support more than regrouping, toward the majority}

Resizing produces larger changes in group support than regrouping in every model (Table~\ref{tab:effects}). Averaged over models, the RMS change is 4.36\pp\ for resize and 2.59\pp\ for regroup, a difference of 1.77\pp\ [1.50, 2.03]. Per model it ranges from 0.79 to 2.94\pp\ with every interval above zero; resize changes are 1.28--2.31 times larger, and larger in 81 of 84 model--topic cells (Figure~\ref{fig:effect}a). The result is robust to how support is aggregated and to the stimulus set: incidence weighting gives 2.53\pp\ [2.26, 2.79] with 84 of 84 cells, and the same direction holds on six further topics (35/36 cells), for five additional models on those topics (30/30), and with a full-answer likelihood readout on two topics (12/12; Appendix Table~\ref{tab:scope}).

The effect of resizing has a clear direction. Its mean signed effect is model-specific, from $-0.34$\pp\ (Phi4-mini) to $+3.19$\pp\ (Gemma4-E2B), but relative to the initial majority of members 1--5 it is consistent: resizing pulls support toward that majority by 1.72\pp\ [1.41, 2.02], with every model's interval above zero (0.61--2.85\pp) and 80 of 84 cells positive. Figure~\ref{fig:effect}b shows the pattern by core composition. With three or four initial supporters among members 1--5, resizing raises support in all six models; with one or two, it lowers support in 8 of 12 model--composition cells. Regrouping, by contrast, changes outcomes by 1.2--4.4\pp\ RMS but has a majority-directed shift of at most 0.03\pp\ in every model (Figure~\ref{fig:effect}c): it reshuffles outcomes configuration by configuration without a systematic direction. Larger meetings act as majority amplifiers, matching majority-following in LLM groups \citep{weng2025conformity,demarzo2026majority}, here with pair contacts held fixed.

\begin{figure*}[t]
\centering
\includegraphics[width=\textwidth]{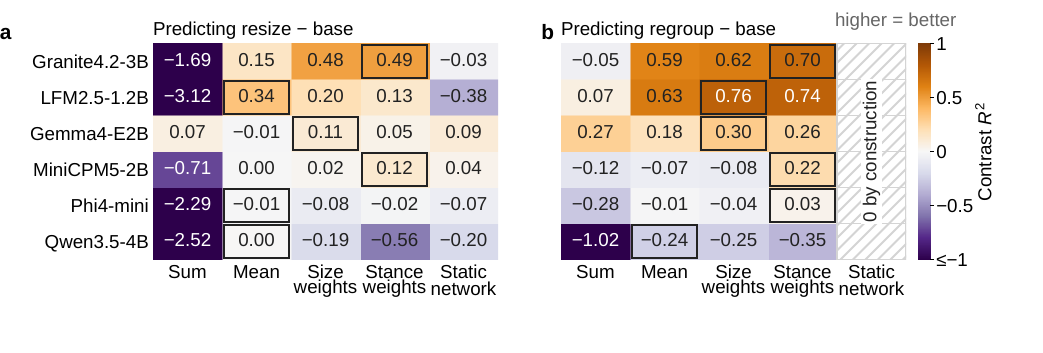}
\caption{Contrast $R^2$ of five pairwise predictors for (a)~resize and (b)~regroup effects (14 topics per model). $R^2=0$ equals predicting no change; colours are clipped at $-1$. Boxes mark each model's best predictor. Sum, Mean, Size and Stance are meeting-wise (Eq.~\ref{eq:z}); Static network applies Size weights to the aggregated contact network (Eq.~\ref{eq:net}) and predicts exactly zero regroup effect.}
\label{fig:pred}
\end{figure*}

\begin{figure*}[t]
\centering
\includegraphics[width=\textwidth]{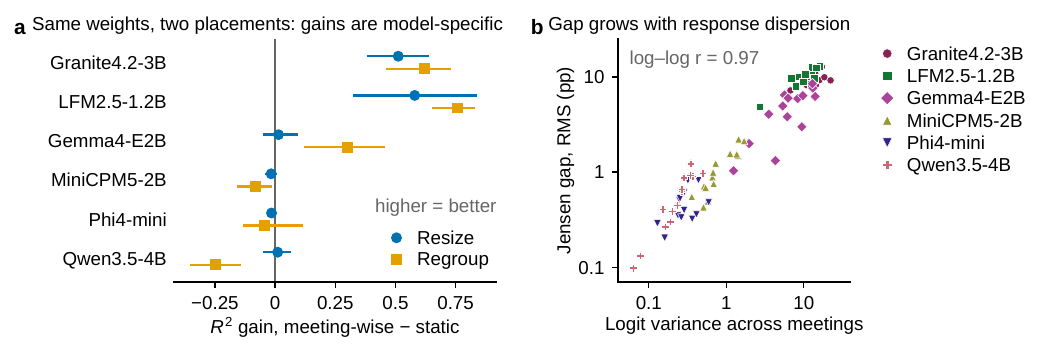}
\caption{Aggregation placement and its mechanism. (a)~Paired $R^2$ gain of meeting-wise Size weights over the same weights on the static contact network, for resize and regroup effects (95\% topic-cluster intervals, 14 topics). (b)~Auxiliary five-person design (three five-person meetings versus all ten triads; every pair meets three times in both): RMS Jensen gap of the Size rule against the mean variance of each member's predicted logits across its meetings (one point per model--topic, $n=84$).}
\label{fig:place}
\end{figure*}

\subsection{Which pairwise rules predict format effects}

Figure~\ref{fig:pred} maps every predictor onto every model. The clearest gains are on regrouping, which the static network cannot see. Meeting-wise Size weights explain 76\% of the regroup effect for LFM2.5-1.2B, 62\% for Granite4.2-3B and 30\% for Gemma4-E2B, and Stance weights reach 70\% for Granite4.2-3B and 22\% for MiniCPM5-2B. For these models, most of the regroup effect is pairwise influence that depends on who shares a meeting, information that a contact count cannot carry.

On resizing, the best rule reaches $R^2=0.49$ for Granite4.2-3B (Stance; Size gives 0.48 [0.37, 0.57]), 0.34 for LFM2.5-1.2B (Mean), and 0.11--0.12 for Gemma4-E2B and MiniCPM5-2B. Influence in larger meetings is strongly sub-additive: summing pairwise effects overshoots, with Sum $R^2$ between $-0.71$ and $-3.12$ for five of six models. The transferred size weights quantify this sub-additivity in the form predicted by social impact theory \citep{latane1981psychology}. Per-peer weights fall from $w_2=0.50$--0.77 to $w_4=0.29$--0.49, so four peers exert 1.16--1.28 times the total influence of two, a sublinear growth with exponent 0.21--0.36 (Appendix Table~\ref{tab:weights}). For Phi4-mini and Qwen3.5-4B, no low-parameter rule beats the no-change baseline on resizing ($R^2\le0.01$); Section~\ref{sec:placement} shows why placement cannot help these two models.

The best weight form is model-specific (Appendix Table~\ref{tab:form}). Size weights beat $1/n$ weights for Granite4.2-3B ($+0.33$ [0.13, 0.51]) and Gemma4-E2B ($+0.12$ [0.06, 0.19]), and stance weights add $0.10$ [0.05, 0.14] for MiniCPM5-2B, while LFM2.5-1.2B and Qwen3.5-4B are best served by the simple mean. As in human conformity research, no single size function fits every setting \citep{bond2005group}. The meeting-wise Size rule also reproduces the majority amplification: it predicts 2.26 of the observed 2.85\pp\ for Granite4.2-3B and 0.65 of 1.72\pp\ overall, against 0.08\pp\ for the static network rule (Table~\ref{tab:effects}).

\subsection{Same weights, different placement}
\label{sec:placement}

Holding the Size weights fixed and changing only where they are aggregated separates two groups of models (Figure~\ref{fig:place}a). For resize effects, meeting-wise computation raises $R^2$ by 0.58 [0.32, 0.84] for LFM2.5-1.2B and 0.51 [0.38, 0.64] for Granite4.2-3B, improving all 14 topics in both; for the other four models the two placements agree to within 0.02. For regroup effects the gain equals the meeting-wise $R^2$, because the network predicts zero: 0.76, 0.62 and 0.30 for LFM2.5-1.2B, Granite4.2-3B and Gemma4-E2B. Where placement helps, the meeting-wise correction points the right way: its correlation with the network's residual is 0.71 (Granite4.2-3B) and 0.75 (LFM2.5-1.2B), and across the 84 model--topic cells the RMS disagreement between the placements correlates 0.77 with the placement gain.

\paragraph{Aggregation error as a Jensen gap.}
By Eqs.~\ref{eq:Q} and \ref{eq:net}, the two placements differ by
\begin{equation}
\hat Q-\hat Q^{\mathrm{net}}=\frac{1}{6}\sum_i J_i,\qquad J_i=\frac{1}{m_i}\sum_{e\ni i}\sigma(z_{ie})-\sigma(\bar z_i),
\label{eq:jensen}
\end{equation}
the Jensen gap of the sigmoid over member $i$'s meetings \citep{ruel1999jensens}. The sigmoid is neither convex nor concave, so $J_i$ can take either sign, and Taylor's theorem bounds it by the dispersion of the meeting-conditional logits,
\begin{equation}
|J_i|\le\frac{\sqrt3}{36}\,\mathrm{Var}_e(z_{ie})
\label{eq:bound}
\end{equation}
(Appendix~\ref{app:jensen}). A contact network is therefore an accurate summary of a pairwise rule exactly when the rule's logits vary little across the meetings a member attends. We test this in an auxiliary design in which five members meet either in three five-person meetings or in all ten triads, with every pair meeting three times in both arms. Each member's five-person meetings are identical, so the placement difference equals the negative triad-side gap; a numerical reconstruction over all 15,552 auxiliary-design conditions matches this identity to within $6.4\times10^{-16}$, and the bound holds in every condition. The logit dispersion predicts the size of the gap almost perfectly (log--log $r=0.97$; Figure~\ref{fig:place}b). It also explains the split between models: Granite4.2-3B, LFM2.5-1.2B and Gemma4-E2B have mean logit variances of 7--14 across meetings, whereas MiniCPM5-2B, Phi4-mini and Qwen3.5-4B have 0.25--0.88, which is why placement makes no difference for the latter three in the main study. Dispersion sets how far the placements can disagree, and alignment with the network's residual decides which is more accurate; both conditions hold for Granite4.2-3B and LFM2.5-1.2B.

The gap also separates two notions that a contact network conflates. With $1/n$ weights, every peer in the auxiliary design receives the same normalised effective weight of $1/4$ in both arms, so the network predicts exactly zero difference. The meeting-wise $1/n$ predictions nonetheless differ by up to 8.46\pp\ RMS (Appendix Table~\ref{tab:simplemean}). Equal effective edge weights do not imply equal predictions once responses pass through a nonlinear readout.

\section{Discussion}

\paragraph{Meeting structure belongs in the simulation state.}
In every model we tested, the same pairwise contacts elicit different support depending on how they are packaged into meetings, and larger meetings consistently amplify the majority. A log of who talked to whom is therefore not a sufficient record of an LLM interaction environment: meeting sizes and co-attendance shape the responses that a sequential simulation carries forward. The contact-matched design is a general attribution tool for this question. It changes one structural factor at a fixed exposure budget and can be reused for other collective outcomes such as polarisation, norm emergence or diffusion.

\paragraph{When contact counts suffice.}
The Jensen bound gives a practical test. Once a pairwise rule has been fitted for a model, the spread of its predicted logits across each member's meetings is computable before any group experiment is run. When the spread is small, as for MiniCPM5-2B, Phi4-mini and Qwen3.5-4B, a contact network loses little relative to the meeting-level rule; when it is large, as for Granite4.2-3B and LFM2.5-1.2B, meeting-wise computation is needed and gains up to 0.58 $R^2$. This parallels two results from network science: temporal and aggregated hypergraph dynamics diverge under nonlinear updates \citep{neuhauser2021temporal}, and exact pairwise reduction of nonlinear social impact requires state-dependent weights \citep{llabres2026reducibility}. Our Stance weights are one such state-dependent family, and they add accuracy where a model's response depends on the stance mix of the room (MiniCPM5-2B on both effects, Granite4.2-3B on regrouping). Because the best rule differs across models, the predictor ladder used here is a template for validating the reduction a simulation relies on \citep{anthis2025llm,larooij2025validation}.

\paragraph{Limitations and next steps.}
The evidence covers six small open-weight models and 14 topics, and every result is a one-step readout from a fixed initial state. The schedules equate pair contacts but not meeting count or tokens. The arguments are human-written, but every agent is an LLM, so the findings describe LLM collectives rather than human groups. The natural next step is to run the contact-matched resize and regroup schedules as sequential meetings with state updates and measure where the collective ends up.

\section{Conclusion}
Holding members, initial stances and every pairwise contact fixed, changing meeting format still changes how LLM collectives respond: resizing moves support more than regrouping and amplifies the initial majority in all six models. Pairwise rules computed meeting by meeting capture regroup effects that a contact network cannot represent and, for two models, predict resizing far better than the same rules on the network. The difference is an exact Jensen gap whose size is set by the dispersion of meeting-conditional predictions, which tells a modeller in advance when contact counts are enough. Meeting format is thus a variable that LLM social simulations should record, test and model.

\section*{Ethical Statement}
The study uses a public corpus of human-written arguments and open-weight models; it involves no human participants and no personal data. Its claims concern LLM collectives. They should not be read as estimates of human group behaviour, and persuasion-oriented uses of meeting design deserve the same scrutiny as other influence techniques.

\bibliography{references}

\appendix

\setcounter{dbltopnumber}{4}
\renewcommand{\dbltopfraction}{0.95}
\renewcommand{\textfraction}{0.05}
\renewcommand{\dblfloatpagefraction}{0.75}
\makeatletter
\setlength{\@dblfptop}{0pt}
\setlength{\@dblfpsep}{8pt plus 2pt minus 2pt}
\makeatother

\section*{Technical Appendix}

\section{Design Details}
\label{app:design}

\paragraph{Topics.} The 14 main-study motions are: prohibit women in combat; abandon the use of school uniform; abolish capital punishment; subsidize space exploration; abolish the right to keep and bear arms; abandon marriage; ban private military companies; abolish intellectual property rights; subsidize vocational education; adopt an austerity regime; the vow of celibacy should be abandoned; legalize prostitution; subsidize journalism; fight for the abolition of nuclear weapons. Six further motions are used for the robustness analyses: legalize sex selection; introduce compulsory voting; fight urbanization; assisted suicide should be a criminal offence; close Guantanamo Bay detention camp; end affirmative action. All 600 arguments (20 motions $\times$ 15 members $\times$ 2 sides) match the ArgKP-2021 train and dev splits verbatim.

\paragraph{Configurations.} Per topic, the 15 pool members form three disjoint five-member cores. Each core is combined with each of the ten members outside it as member 6 and with all 64 stance patterns, giving $3\times10\times64=1{,}920$ configurations per model--topic. The main study comprises $6\times14\times1{,}920=161{,}280$ matched configurations; the released condition-level table, which includes the robustness stages, has 311,040 rows.

\begin{table*}[!tp]
\centering
\small
\begin{tabular}{llll}
\toprule
Label & Hugging Face repository & Revision & Role \\
\midrule
Gemma4-E2B & \texttt{google/gemma-4-E2B-it} & \texttt{3e22461f65} & main + micro \\
Granite4.2-3B & \texttt{ibm-granite/granite-4.2-3b} & \texttt{4e9c275e23} & main + micro \\
LFM2.5-1.2B & \texttt{LiquidAI/LFM2.5-1.2B-Instruct} & \texttt{0f604ada3f} & main + micro \\
MiniCPM5-2B & \texttt{openbmb/MiniCPM5-2B} & \texttt{f97400052a} & main + micro \\
Phi4-mini & \texttt{microsoft/Phi-4-mini-instruct} & \texttt{cfbefacb99} & main + micro \\
Qwen3.5-4B & \texttt{Qwen/Qwen3.5-4B} & \texttt{851bf6e806} & main + micro \\
Gemma4-E4B & \texttt{google/gemma-4-E4B-it} & \texttt{ee0ef60236} & micro; earlier topics \\
Ministral3-3B & \texttt{mistralai/Ministral-3-3B-Instruct-2512-BF16} & \texttt{b6d637bef2} & micro; earlier topics \\
Nemotron3-Nano-4B & \texttt{nvidia/NVIDIA-Nemotron-3-Nano-4B-BF16} & \texttt{dfaf35de3e} & micro; earlier topics \\
Qwen3.5-2B & \texttt{Qwen/Qwen3.5-2B} & \texttt{15852e8c16} & micro; earlier topics \\
SmolLM3-3B & \texttt{HuggingFaceTB/SmolLM3-3B} & \texttt{a07cc9a04f} & micro; earlier topics \\
 
\bottomrule
\end{tabular}
\caption{Models and pinned download revisions (first ten hex digits). ``Main + micro'': main group study and micro-experiment; the other five models enter the micro-experiment and the robustness analysis on the six further topics.}
\label{tab:models}
\end{table*}

\begin{table*}[!tp]
\centering
\small
\setlength{\tabcolsep}{3pt}
\begin{tabular}{llccc}
\toprule
Scope & Weight & Res.\ $-$ reg. & Cells & Models \\
\midrule
Main: 6 models, 14 topics & person & 1.77 [1.50, 2.03] & 81/84 & 6/6 \\
 & incidence & 2.53 [2.26, 2.79] & 84/84 & 6/6 \\
6 models, 6 further topics & person & 1.68 [1.41, 1.97] & 35/36 & 6/6 \\
 & incidence & 2.45 [2.19, 2.69] & 36/36 & 6/6 \\
5 other models, 6 further topics & person & 1.30 [1.22, 1.37] & 30/30 & 5/5 \\
 & incidence & 1.79 [1.69, 1.89] & 30/30 & 5/5 \\
6 models, 2 topics, R1 & person & 1.68 & 12/12 & 6/6 \\
 & incidence & 2.37 & 12/12 & 6/6 \\
 
\bottomrule
\end{tabular}
\caption{RMS difference between resize and regroup (pp, [95\% CI]) across study scopes and weightings. The R1 readout scores full candidate answers by log-likelihood; with two topics, no cluster-bootstrap interval is reported.}
\label{tab:scope}
\end{table*}

\section{Predictor Details}
\label{app:pred}

Table~\ref{tab:r2full} gives the intervals behind Figure~\ref{fig:pred}, Table~\ref{tab:form} the paired weight-form comparisons, Table~\ref{tab:weights} the transferred Size weights, and Table~\ref{tab:majpred} the majority-directed shift implied by each rule.

\begin{table*}[!tp]
\centering
\small
\setlength{\tabcolsep}{3pt}
\begin{tabular}{lccccc}
\toprule
Model & Sum & Mean & Size weights & Stance weights & Static network \\
\midrule
\multicolumn{6}{l}{\textit{Resize $-$ base}} \\
Granite4.2-3B & $-$1.69 [$-$2.68, $-$0.88] & 0.15 [0.01, 0.29] & 0.48 [0.37, 0.57] & 0.49 [0.39, 0.59] & $-$0.03 [$-$0.20, 0.12] \\
LFM2.5-1.2B & $-$3.12 [$-$4.97, $-$1.58] & 0.34 [0.21, 0.45] & 0.20 [0.02, 0.37] & 0.13 [$-$0.04, 0.27] & $-$0.38 [$-$0.55, $-$0.21] \\
Gemma4-E2B & 0.07 [$-$0.10, 0.19] & $-$0.01 [$-$0.05, 0.03] & 0.11 [0.05, 0.19] & 0.05 [$-$0.02, 0.12] & 0.09 [0.03, 0.15] \\
MiniCPM5-2B & $-$0.71 [$-$1.81, $-$0.27] & 0.00 [$-$0.02, 0.02] & 0.02 [$-$0.10, 0.11] & 0.12 [$-$0.01, 0.21] & 0.04 [$-$0.09, 0.13] \\
Phi4-mini & $-$2.29 [$-$4.91, $-$0.83] & $-$0.01 [$-$0.02, 0.01] & $-$0.08 [$-$0.23, 0.07] & $-$0.02 [$-$0.12, 0.09] & $-$0.07 [$-$0.22, 0.07] \\
Qwen3.5-4B & $-$2.52 [$-$4.69, $-$1.05] & 0.00 [$-$0.06, 0.05] & $-$0.19 [$-$0.34, $-$0.03] & $-$0.56 [$-$0.83, $-$0.37] & $-$0.20 [$-$0.32, $-$0.07] \\
\midrule
\multicolumn{6}{l}{\textit{Regroup $-$ base}} \\
Granite4.2-3B & $-$0.05 [$-$0.63, 0.31] & 0.59 [0.46, 0.69] & 0.62 [0.46, 0.73] & 0.70 [0.60, 0.78] & 0 (constr.) \\
LFM2.5-1.2B & 0.07 [$-$0.17, 0.29] & 0.63 [0.57, 0.69] & 0.76 [0.65, 0.83] & 0.74 [0.64, 0.82] & 0 (constr.) \\
Gemma4-E2B & 0.27 [0.07, 0.48] & 0.18 [0.03, 0.29] & 0.30 [0.12, 0.46] & 0.26 [0.05, 0.45] & 0 (constr.) \\
MiniCPM5-2B & $-$0.12 [$-$0.29, 0.02] & $-$0.07 [$-$0.12, $-$0.01] & $-$0.08 [$-$0.16, $-$0.01] & 0.22 [0.04, 0.36] & 0 (constr.) \\
Phi4-mini & $-$0.28 [$-$0.42, $-$0.09] & $-$0.01 [$-$0.08, 0.10] & $-$0.04 [$-$0.13, 0.11] & 0.03 [$-$0.08, 0.16] & 0 (constr.) \\
Qwen3.5-4B & $-$1.02 [$-$1.37, $-$0.66] & $-$0.24 [$-$0.34, $-$0.14] & $-$0.25 [$-$0.35, $-$0.14] & $-$0.35 [$-$0.88, $-$0.01] & 0 (constr.) \\
 
\bottomrule
\end{tabular}
\caption{Contrast $R^2$ [95\% CI] for every predictor, model and intervention (values shown in Figure~\ref{fig:pred}). ``0 (constr.)'': identically zero because base and regroup share the contact network.}
\label{tab:r2full}
\end{table*}

\begin{table*}[!tp]
\centering
\begin{minipage}[t]{\columnwidth}
\vspace{0pt}
\centering
\small
\setlength{\tabcolsep}{3.5pt}
\begin{tabular}{lcc}
\toprule
Model & Size $-$ Mean & Stance $-$ Size \\
\midrule
Granite4.2-3B & 0.33 [0.13, 0.51] & 0.01 [$-$0.09, 0.10] \\
LFM2.5-1.2B & $-$0.14 [$-$0.33, 0.04] & $-$0.07 [$-$0.18, 0.04] \\
Gemma4-E2B & 0.12 [0.06, 0.19] & $-$0.06 [$-$0.10, $-$0.02] \\
MiniCPM5-2B & 0.02 [$-$0.11, 0.12] & 0.10 [0.05, 0.14] \\
Phi4-mini & $-$0.07 [$-$0.22, 0.07] & 0.07 [$-$0.01, 0.14] \\
Qwen3.5-4B & $-$0.19 [$-$0.31, $-$0.07] & $-$0.38 [$-$0.61, $-$0.21] \\
 
\bottomrule
\end{tabular}
\caption{Weight-form comparisons on resize effects: paired $R^2$ gains [95\% CI], both meeting-wise. Positive values favour the first rule.}
\label{tab:form}
\end{minipage}\hfill
\begin{minipage}[t]{\columnwidth}
\vspace{0pt}
\centering
\small
\begin{tabular}{lccc}
\toprule
Model & $w_2$ & $w_4$ & $4w_4/2w_2$ \\
\midrule
Granite4.2-3B & 0.662 & 0.403 & 1.22 \\
LFM2.5-1.2B & 0.621 & 0.369 & 1.19 \\
Gemma4-E2B & 0.766 & 0.490 & 1.28 \\
MiniCPM5-2B & 0.598 & 0.373 & 1.24 \\
Phi4-mini & 0.622 & 0.359 & 1.16 \\
Qwen3.5-4B & 0.504 & 0.294 & 1.17 \\
 
\bottomrule
\end{tabular}
\caption{Transferred Size weights (mean over the 14 leave-topic-out fits). The last column is the total influence of four peers relative to two.}
\label{tab:weights}
\end{minipage}
\end{table*}

\section{Nonlinear Averaging}
\label{app:jensen}

\paragraph{Bound.} With $\bar z_i$ the mean of member $i$'s meeting logits, Taylor's theorem gives $\sigma(z)=\sigma(\bar z_i)+\sigma'(\bar z_i)(z-\bar z_i)+\tfrac12\sigma''(\xi)(z-\bar z_i)^2$. Averaging over meetings removes the linear term, so $|J_i|\le\tfrac12\sup|\sigma''|\,\mathrm{Var}_e(z_{ie})$ with $\sup|\sigma''|=\sqrt3/18$. Equal logits across meetings are sufficient for $J_i=0$ but not necessary, because curvature on either side of zero can cancel: for logits $-2$ and $2$, the mean of the sigmoids equals $\sigma(0)$.

\paragraph{Auxiliary design.} Five members meet either three times as a group of five or once in each of the ten triads (six triads per member); every pair meets three times in both arms. Under Size weights, a member's three five-person meetings have identical logits, so $\hat\Delta-\hat\Delta^{\mathrm{net}}=-\tfrac15\sum_i J^{\mathrm{tri}}_i$. Under $1/n$ weights the averaged logit is $a_i+\tfrac14\sum_j v_{ij}$ in both arms.

\begin{table*}[!tp]
\centering
\begin{minipage}[t]{\columnwidth}
\vspace{0pt}
\centering
\small
\setlength{\tabcolsep}{3.2pt}
\begin{tabular}{lcccccc}
\toprule
Model & Obs. & Sum & Mean & Size & Stance & Net. \\
\midrule
Granite4.2-3B & 2.85 & $-$0.08 & 0.34 & 2.26 & 2.00 & 0.62 \\
LFM2.5-1.2B & 0.93 & 0.44 & 0.50 & 1.11 & 0.53 & $-$0.16 \\
Gemma4-E2B & 2.27 & 1.15 & 0.06 & 0.68 & 0.38 & 0.25 \\
MiniCPM5-2B & 2.74 & $-$0.78 & 0.06 & $-$0.09 & 0.21 & $-$0.11 \\
Phi4-mini & 0.90 & $-$0.27 & 0.03 & $-$0.01 & 0.02 & $-$0.04 \\
Qwen3.5-4B & 0.61 & $-$0.89 & $-$0.03 & $-$0.04 & $-$0.24 & $-$0.09 \\
\midrule
All six & 1.72 & $-$0.07 & 0.16 & 0.65 & 0.48 & 0.08 \\
 
\bottomrule
\end{tabular}
\caption{Majority-directed resize shift $D$ (pp): observed and predicted by each rule.}
\label{tab:majpred}
\end{minipage}\hfill
\begin{minipage}[t]{\columnwidth}
\vspace{0pt}
\centering
\small
\begin{tabular}{lccc}
\toprule
Model & Observed & Meeting-wise & Network \\
\midrule
Granite4.2-3B & 18.30 & 6.58 & 0.00 \\
LFM2.5-1.2B & 9.52 & 8.46 & 0.00 \\
Gemma4-E2B & 21.48 & 3.02 & 0.00 \\
MiniCPM5-2B & 16.94 & 0.90 & 0.00 \\
Phi4-mini & 10.34 & 0.33 & 0.00 \\
Qwen3.5-4B & 7.12 & 0.67 & 0.00 \\
 
\bottomrule
\end{tabular}
\caption{Auxiliary design with $1/n$ weights, main-study topics (1,344 conditions per model): RMS of the five-versus-triads difference (pp). The network column gives the maximum absolute value, which is zero to machine precision.}
\label{tab:simplemean}
\end{minipage}
\end{table*}

\section{Micro-Experiment}
\label{app:micro}

\begin{figure*}[!tp]
\centering
\includegraphics[width=\textwidth]{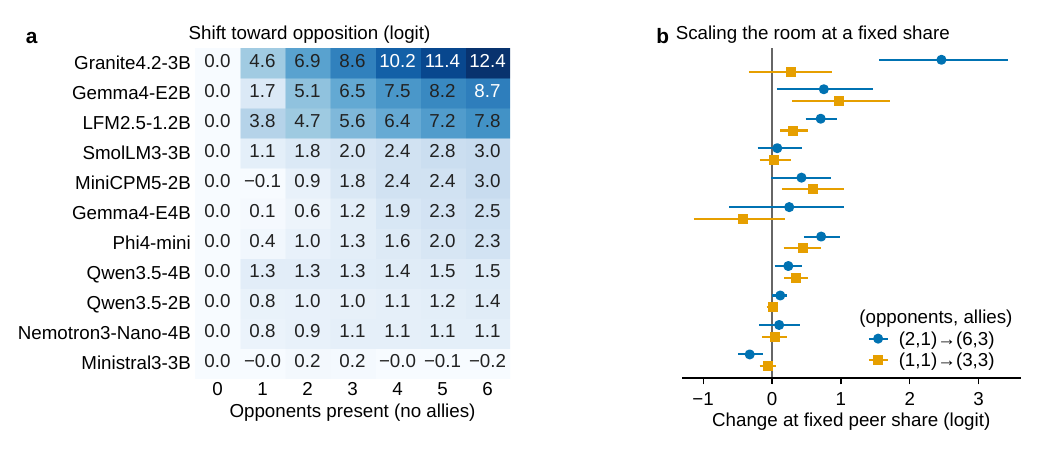}
\caption{Micro-experiment (64 items, 19 topics, 11 models). (a)~Mean change in opposition-directed logit with $k$ opponents and no allies. (b)~Change when peers are scaled at fixed opponent:ally ratios (95\% topic-cluster intervals); rows follow the order in (a).}
\label{fig:micro}
\end{figure*}

A focal member with a fixed stance sees the arguments of 0--6 opponents and 0--3 allies; the outcome is the change in its opposition-directed logit relative to no peers. The experiment covers 11 models and 64 items from 19 topics; the target topic is excluded when weights are transferred to the group study. Figure~\ref{fig:micro}a shows a model-specific count response: six opponents shift the focal logit by 1.13--12.40 in ten models, while Ministral3-3B stays near zero ($-0.20$). In Figure~\ref{fig:micro}b, peers are added at a fixed opponent:ally ratio. At 2:1, scaling from (2,1) to (6,3) still moves several models (Granite4.2-3B: 2.46 [1.56, 3.42] logits), and a 44-item subset gives the same sign for all 11 models. Because the focal member counts as an ally, the room-level opponent share in this contrast rises from 2/4 to 6/10.

\begin{table*}[!tp]
\centering
\small
\begin{tabular}{lcccc}
\toprule
Model & Shift at 6 opponents & Power exponent (1--6) & Neutral-statement effect & Fixed share 2:1 \\
\midrule
Granite4.2-3B & 12.40 & 0.56 [0.47, 0.70] & $-$2.21 [$-$2.80, $-$1.64] & 2.46 [1.56, 3.42] \\
Gemma4-E2B & 8.72 & 0.87 [0.73, 1.09] & 0.34 [0.07, 0.57] & 0.75 [0.07, 1.47] \\
LFM2.5-1.2B & 7.80 & 0.41 [0.36, 0.48] & $-$2.40 [$-$3.04, $-$1.87] & 0.71 [0.49, 0.94] \\
SmolLM3-3B & 3.02 & 0.54 [0.46, 0.63] & $-$0.34 [$-$0.49, $-$0.19] & 0.08 [$-$0.21, 0.44] \\
MiniCPM5-2B & 3.01 & -- & 0.25 [$-$0.22, 0.75] & 0.43 [$-$0.01, 0.86] \\
Gemma4-E4B & 2.49 & 1.77 & $-$0.44 [$-$1.12, 0.21] & 0.25 [$-$0.62, 1.04] \\
Phi4-mini & 2.34 & 0.97 & 0.10 [$-$0.15, 0.36] & 0.71 [0.46, 0.99] \\
Qwen3.5-4B & 1.53 & 0.10 [$-$0.01, 0.18] & $-$0.68 [$-$0.87, $-$0.46] & 0.24 [0.04, 0.44] \\
Qwen3.5-2B & 1.44 & 0.27 [0.19, 0.34] & $-$0.08 [$-$0.14, $-$0.02] & 0.12 [0.00, 0.22] \\
Nemotron3-Nano-4B & 1.13 & 0.24 [0.11, 0.37] & $-$0.41 [$-$0.56, $-$0.26] & 0.10 [$-$0.18, 0.40] \\
Ministral3-3B & $-$0.20 & -- & 0.15 [$-$0.02, 0.30] & $-$0.32 [$-$0.49, $-$0.13] \\
 
\bottomrule
\end{tabular}
\caption{Micro-experiment summaries (logits, [95\% CI]). Power exponent: fitted exponent of the no-ally count curve; ``--'' or no interval where it is undefined. Neutral-statement effect: change in a single opponent's influence when three neutral statements are added.}
\label{tab:micro}
\end{table*}

\section{Reproducibility}
\label{app:process}

{\raggedright\noindent Code repository: \repositoryurl{https://github.com/ZachGu-00/llm-meeting-format}.\par}

The format analysis is an exploratory analysis of saved measurements from a larger project; intervals are pointwise, from a topic-cluster bootstrap with 20,000 draws and seed 20261105. The supplementary material contains the condition-level tables, the transferred weights, all derived statistics and the scripts that generate every table and figure, together with the project's earlier confirmatory tests and a generated-dialogue study, which address different hypotheses and are documented there in full.

\end{document}